\documentstyle[preprint,aps,prd]{revtex}
\begin{document}
\thispagestyle{empty}
{\baselineskip0pt
\leftline{\large\baselineskip16pt\sl\vbox to0pt{\hbox{Yukawa Institute Kyoto}
               \hbox{ }\vss}}
\rightline{\large\baselineskip16pt\rm\vbox to20pt{\hbox{YITP/K1103}
           \hbox{KUNS-1328}
           \hbox{astro-ph/9507085}
           \hbox{July 1995}
\vss}}%
}
\vskip3cm
\begin{center}{\large
Volume Expansion Rate and The Age of The Universe}
\end{center}
\begin{center}
{\large Takashi Nakamura, $^{\dag}$Ken-ichi Nakao, $^{\dag}$Takeshi Chiba} \\
{\large and $^{\dag}$Tetsuya Shiromizu} \\
{\em Yukawa Institute for Theoretical Physics,~Kyoto University} \\
{\em Kyoto 606-01,~Japan}\\
{\em $^{\dag}$Department of Physics,~Kyoto University} \\
{\em Kyoto 606-01,~Japan}
\end{center}
\begin{abstract}
Under four assumptions such that
1) Einstein's theory of gravity is
correct, 2) existence of foliation by geodesic slicing,
3) the trace of the extrinsic curvature, $K$, is
negative at the present time in observed region,
that is, the observed universe is now expanding,
4) the strong energy condition is satisfied, we show that
$H_{v0} t_{0} \leq 1$, where $H_v \equiv - K/3$ agrees
with the Hubble parameter in the case of a
homogeneous and isotropic universe,
and $t_0$ is the age of the Universe.
If $H_{v0} t_0 > 1$ is confirmed observationally, at least one of the
four assumptions is incorrect.


\end{abstract}

\newpage

Recent observations of Cepheid variables in NGC4571 (Pierce et al. 1994)
and M100 (Freedman et al. 1994) suggest that the Hubble parameter $(H_0)$ is
$\sim$ 80km/s/Mpc (Fukugita et al. 1993; Jacoby et al. 1992;
van den Bergh 1992).
Such a high value of $H_0$ may contradict the age estimate of
our universe
using, for example, globular clusters (Demanqul et al. 1991; Renzini 1991)
so that $H_0 t_0 > 1$.
However one may consider an inhomogeneous universe such that $H_0$ in our
neighborhood is high (Turner et al. 1992) but global $H_0$ is
low enough to agree with $t_0$.
In the previous paper (Nakao et al. 1995) we considered a simple inhomogeneous
model
in which we are in a void expressed by an open Friedmann universe and showed
that $H_0 t_0 \leq 1$ even in such an inhomogeneous model as long as
the peculiar velocity correction is perfect.
In this paper, we extend our argument to a more general inhomogeneous
universe.

We shall put four assumptions on the evolution of the universe.

{\bf 1}) First we assume that Einstein's theory of gravity is the correct
theory
to describe the evolution of the universe after the Planck time.
Then the basic equations
in (3+1)-formalism become: the constraint equations are given by
\begin{eqnarray}
^{(3)}R + K^2  &=&K_{i j}~K^{ij} + 16 \pi \rho_{H}, \\
 K^j_{i | j} -  K_{| i}  &= &8 \pi{J}_i ,
\end{eqnarray}
the evolution equations
\begin{eqnarray}
\frac{\partial K_{i}^{j}}{\partial t}
& = &\alpha (^{(3)}R_{i}^{j} +K K_{i}^{j})
  - {\alpha_{| i}}^{|j}             \nonumber \\
& - &8 \pi\Bigl(S_{i}^{j}
  + \frac{1}{2} \delta_{i}^{j}(\rho_H - S^l_l)\Bigr)  \nonumber \\
& - &K^{m}_{i} \beta^{j}_{|m} + K_m^j \beta^{m}_{|i}
  + K_{i|m}^{j}\beta^m, \\
\frac{\partial \gamma_{i j}}{\partial t}
& =&
-2 \alpha K_{i j} + \beta_{i|j} + \beta_{j|i},
\end{eqnarray}
where $\alpha$ and $\beta_i$ are the lapse function and the shift vector,
respectively, and the vertical bar means the covariant derivative
with respect to the 3-metric $\gamma_{ij}$.
$\rho_{H}$, $ J_i$ and $S_{i j}$ are defined by
\begin{eqnarray}
\rho_{H} & =&  T^{\mu \nu}n_{\mu}n_{\nu}, \\
J_i      & =&  -T^{\mu \nu}n_{\mu} h_{i \mu}, \\
S_{i j}  & =&  T^{\mu \nu}h_{i \mu} h_{j \nu},
\end{eqnarray}
and
\begin{equation}
h_{\mu \nu} =  g_{\mu \nu} + n_\mu n_\nu,
\end{equation}
where $n^{\mu}$ and $T^{\mu \nu}$ are the normal vector
to $t =$constant hypersurface and the energy momentum tensor
of the matter, respectively.

{\bf 2}) We assume the existence of a foliation by
geodesic slicing, (i.e, $\alpha = 1$) at least in the region
over which we can, in principle, observe.\footnote{{ \em
The region over which we can, in principle, observe} is described
mathematically as the causal past $J^{-}(p)$ of our present
world point denoted by $p$.}
Further for simplicity we choose $\beta_i = 0$
so that the line element can be expressed as
\begin{equation}
ds^2  = -dt^2 + \gamma_{i j}~dx^i~dx^j.
\end{equation}

As for the existence of the foliation by the
geodesic slices, see Appendix A.

{\bf 3}) The observed universe is expanding, i.e.,
$K <0$.\footnote{In mathematical language,
the signature of $K$ is negative in the intersection of each
geodesic slice with our observed region, $O_p \cap J^{-}(p)$,
where $O_{p}$ is the neighborhood of $p$.}

{\bf 4}) The strong energy condition is satisfied, i.e.,
$S_{l}^{l}+\rho_{H} \geq 0$.

Now from the trace of Eq.(4), we have
\begin{equation}
K = -\frac{1}{\sqrt \gamma}\frac{\partial {\sqrt \gamma}}{\partial t},
\end{equation}
where $\gamma$ is the determinant of $\gamma_{i j}$.
We shall call $-K$ the volume expansion rate and define $H_v$ as
\begin{equation}
 H_v(t, x) = -\frac{K(t,x)}{3}.
\end{equation}
For a homogeneous and isotropic universe, $H_{v}$
agrees with the Hubble parameter.
The trace of Eq.(3) with the aid of the Hamiltonian constraint
equation, Eq.(1), yields
\begin{equation}
\frac{\partial K}{\partial t} = \frac{1}{3}~K^2
+(K^{T}_{i j})^{2}+4\pi (S_{l}^{l}+\rho_{H}),
\end{equation}
where $K_{ij}^{T}$ is the traceless part of $K_{ij}$.
 From the assumption {\bf 4)}, Eq.(12) implies that $K$ is a monotonically
increasing function with respect to $t$ and hence
from the assumptions {\bf 2}) and {\bf 3}), $K$ is always negative
in ``our past".\footnote{Here ``our past" means the subset of
$J^{-}(p)$ which is connected to our observed region by the
timelike geodesic $n^{\mu}$ perpendicular to the geodesic slices.}
Dividing Eq.(12) by $K^{2}$, we have
\begin{equation}
\frac{1}{K^2}\frac{\partial K}{\partial t}
=\frac{1}{3} +  f,
\end{equation}
where
\begin{equation}
f  = {1 \over K^{2}}\Bigl((K_{ij}^{T})^{2}
+ 4\pi(S_{l}^{l}+\rho_{H})\Bigr) \geq 0,
\end{equation}
from the assumptions {\bf 4}).

Integrating Eq.(13) from $t = t_i$ to  $t_0$, we have
\begin{eqnarray}
 -\frac{1}{K_{(0)}}+\frac{1}{K_{(i)}}
& = &\frac{1}{3}(t_{0}-t_{i})
+  \int_{t_{i}}^{t_0}fdt  \nonumber \\
&\geq &\frac{1}{3}(t_{0}-t_{i}),
\end{eqnarray}
where $K_{(0)}=K(t_{0},x)$ and $K_{(i)}=K(t_{i},x)$.
Since $K$ is always negative, $H_{vi}=-K_{(i)}/3>0$, and hence
using expression (11), we can rewrite Eq.(15) as
\begin{equation}
H_{v0}(t_0-t_{i})  \leq 1-\frac{H_{v0}}{H_{vi}} \leq 1,
\end{equation}
where $H_{v0}=-K_{(0)}/3$.
Choosing $t_{i}$ to be at the initial
singularity of the universe $t=0$,
we obtain $H_{v0}t_{0} \leq 1$.

However, it should be noted that $t_{i}$ can be of course freely chosen.
We can choose $t_{i}$ to be the equal time $t_{eq}$ or to be the
decoupling time $t_{dec}$ of our universe. The important fact shown here
is just that the period from arbitrary $t_{i}$ to the present time $t_{0}$
is bounded by $H_{v0}^{-1}$ (see Appendix B for the more detailed argument).

Equation (16) is a local equation.
If $H_{v0} t_{0} > 1$ is confirmed observationally in our neighborhood,
at least one of the four assumptions which lead to Eq.(16) is incorrect.
Four possibilities are:

\begin{enumerate}
\item Einstein's gravity is not correct.
\item The foliation by geodesic slices does not exist (on any averaging scale
smaller than our observed scale; see Appendix A).
\item The observed universe is now not expanding.
\item The strong energy condition is not satisfied.
\end{enumerate}

To include the cosmological constant belongs to the fourth possibility.
However we note that this is not the only one. We will discuss each possibility
elsewhere (Nakamura et al. 1995).

Since $H_{v0}$ is defined by the volume expansion rate (Eq.(11)),
we should consider the relation between the observed Hubble parameter
$H_{0}$ and $H_{v0}$. In order to do so, we need to investigate the
null geodesic $k^{\mu}$ and the distance-redshift relation.
By virtue of the comoving coordinate Eq.(9),
the angular frequency of the light ray
for the comoving observer and comoving source
(perhaps the cluster of galaxy) is given simply by
$\omega=k^{t}$ and hence $k^{\mu}=(\omega,k^{i})$.

It is sufficient for our purpose to see
the time-component of the geodesic equation,
\begin{equation}
{d\omega \over d\lambda}=K_{ij}k^{i}k^{j},
\end{equation}
and the null condition which is given by
\begin{equation}
\omega={d\ell \over d\lambda}\equiv \sqrt{\gamma_{ij}k^{i}k^{j}}.
\end{equation}
Integrating Eq.(17), we obtain
\begin{equation}
\omega_{0}=\omega_{e}
+\int_{\lambda_{e}}^{\lambda_{0}}K_{ij}k^{i}k^{j}d\lambda,
\end{equation}
where $\omega_{0}$ and $\omega_{e}$ are respectively the observed
angular frequency and emitted one.
Here we assume that the proper distance $\Delta\ell$
between the observer and the source
is so small that the integrand in Eq.(19) does not change rapidly.
This condition may correspond to $|K_{ij}|\Delta\ell \ll 1$ (or
$|K_{ij}|\omega_{e}\Delta\lambda \ll 1$,
where $\Delta\lambda=\lambda_{0}-\lambda_{e}$,
because  $\Delta\ell \sim \omega_{e}\Delta\lambda$ by Eq.(18)).
Then from the above equation, the redshift $z$ is approximately written as
\begin{equation}
z={\omega_{e} \over \omega_{0}}-1 \sim
 -{1\over \omega_{o}}\Bigl(K^{T}_{ij}+{1\over3}\gamma_{ij}K\Bigr)
 k^{i}k^{j}\Delta\lambda.
\end{equation}
Using Eq.(18), we get the Hubble law in an inhomogeneous
universe as
\begin{equation}
     z \sim H_{v0}\Delta\ell
       -K^{T}_{ij}\frac{k^{i}k^{j}}{\omega_{o}^{2}}\Delta\ell.
\end{equation}
In general, $K^{T}_{ij}$ does not vanish. However, if the
observer stays in an almost isotropic region with the linear
Hubble law $z \sim H_{0}\Delta \ell$,
the second term of the R.H.S. in the above equation is
much smaller than the first term near the observer:
If the observer stays in the almost isotropic region,
such a situation can be approximated by a Tolman-Bondi
space-time in which the observer is at the symmetric center.
In this space-time, denoting the comoving radial coordinate
by $r$, $K^{T}_{ij}k^{i}k^{j} \propto r^{2} \propto \Delta\ell^{2}$
near the observer. This means that if the observer can
find the effect of the second term in Eq.(21), the Hubble law
is not linear. Conversely, we can ignore the second term in Eq.(21) in
the situation in which the Hubble law is linear.

Lauer and Postman (1994) suggested from the observations of brightest
cluster galaxies over $0.01 \leq z \leq 0.05$ that
the Hubble flow is essentially uniform and isotropic.
In this case we may regard $H_{v0} \sim H_0$.
So ${H}_{0}t_0 > 1$ requires the four possibilities stated above even if
our universe is so inhomogeneous globally that $H_{v}$ in other places
with distance greater than $100$Mpc is much smaller than
80km/s/Mpc.

We would like to thank M. Fukugita for useful lectures
on the Hubble parameter and T. Tanaka and M. Siino
for their useful discussion. We are grateful to S.A. Hayward for
his careful reading of the English.
This work was supported by  Grant-in-Aid Scientific Research
of the Ministry of Education 04234104 and Grant-in-Aid
for Scientific Research Fellowship 2925.

\appendix
\section*{A}

The geodesic slicing condition imposes the hypersurface unit normal
$n^{\mu}$ to be tangent to timelike geodesics.
Here it should be noted that the nearby timelike geodesics
have a tendency to cross with each other
in region curved by gravity, e.g. due to the concentration of matter.
If the crossing of the timelike geodesics
with tangent $n^{\mu}$ occurs, $n^{\mu}$ becomes multi-valued at this
crossing point. This means that the hypersurface becomes singular at this
point since the normal vector and the normal direction to the
hypersurface can not be defined uniquely there.
Hence a foliation by regular geodesic slices through this
crossing point does not exist.
If we consider the galaxy, star or much smaller object, e.g., a stone,
the timelike geodesics through those objects
may cross on the free fall time determined by the
energy density of the object (Smarr $\&$ York 1978). Hence, rigorously
speaking,
the foliation by geodesic slicing beyond the shortest free fall
time scale of the system may not exist.

However here it should be noted that the existence of the foliation by
geodesic slices in an approximate sense depends on what scale we consider.
When we investigate, for example, the formation of a star,
we consider the matter averaged over an appropriate scale so that the matter
can be treated as a continuous quantity (the metric tensor
correspondingly becomes an averaged one).
In such a treatment the geodesic slicing may be
applicable during {\em the free fall time determined by the averaged
density of the star in the above sense, not the free fall time determined by
the
density of the nuclei of the atoms themselves} and we know that
such an averaged treatment well describes the dynamics of the star.

Here we are considering the averaged matter and metric tensor in a
cosmological sense: the averaging scale should be determined
so that the free fall time scale agrees with the age of the universe.
It is usually considered that under such an averaging treatment
a bound object such as a cluster of galaxies can be regarded as
a particle which follows the geodesic $n^{\mu}$, assuming that the
rotation of the velocity field associated with the cluster of galaxies
is negligible.

\appendix
\section*{B}

The result Eqs.(15) and (16) is essentially the same as
the well known fact that within the proper time $\tau \leq H_{v0}^{-1}$
measured toward the past there exists a conjugate point to a
hypersurface $\Sigma$
with $K_{(0)}=-3H_{v0}$ (Hawking $\&$ Ellis 1973; Wald 1984).
However it is worthy to note that under the assumptions
{\bf 1})$\sim${\bf 4}) the occurrence of the
conjugate point to $\Sigma$ means the existence of singularities by
almost the same argument as Wald's one (Wald 1984).

Before we proceed to our discussion, we shall see briefly
the Wald's singularity theorem (Wald 1984). Assuming the
Einstein equations and
\begin{description}
\item[A)] the space-time is globally hyperbolic,
\item[B)] the strong energy condition holds,
\item[C)] there exists a Cauchy surface $\Sigma$ for which the trace
of the extrinsic curvature satisfies $K \leq -3H_{v0} <0$ everywhere,
\end{description}
then $H_{v0}t_0 \leq 1$, where $t_0$ is the proper time of all past
directed timelike curves from $\Sigma$.

The proof is very simple. {\em If there exists a timelike curve which
is longer than $H_{v0}^{-1}$ from $\Sigma$, there exists a maximal
length curve which is longer than $H_{v0}^{-1}$ and has no conjugate points
from the global hyperbolicity.
This implies a contradiction with the conditions {\bf A)}
and {\bf C)} because these conditions imply that all
timelike geodesics have conjugate points.}

 From the above theorem, we can read that the age of the universe is
bounded by $H_{v0}^{-1}$. However it should be noted that
the above conditions are global statements.
Even if we observe the universe perfectly,
we can not, in principle, confirm whether the above conditions are
satisfied.
Thus global assumptions in the above theorem are not suitable for
the present problem. In this paper, our
discussion is restricted only in the observed and observable region
and we consider what we can say from the
observation on $H_{0}$.

Here, we shall show the existence of the singularity in the past
under our assumptions {\bf 1})$\sim${\bf 4}) although it
seems to be trivial.
{\em Suppose that the universe $J^{-}(p)$ is foliated past-completely
by geodesic slicing beyond the time $t-t_{i}>H^{-1}_{v0}$, that is,
the age of the universe is longer than $H^{-1}_{v0}$.
On the other hand, the foliation by geodesic slicing must break down
within $t-t_{i}>H^{-1}_{v0}$ because we see from Eq.(15)
that $K_{(i)} < 3[t_{0}-t_{i}-H_{v0}^{-1}]^{-1}  \rightarrow -\infty$
for $t_{0}-t_{i} \rightarrow H^{-1}_{v0}$. This contradicts
the past-complete foliation by geodesic slices. Hence there
exists a past incomplete timelike geodesic $n^{\mu}$
perpendicular to the
geodesic slices. Especially, the proper time of the timelike
geodesic $n^{\mu}$ in our past ({\rm see footnote}) is bounded by
$H^{-1}_{v0}$.}
Hence, the assumptions {\bf 1)}$\sim${\bf 4)}
represents a kind of Big Bang cosmology.


\vskip 1cm
\begin{centerline}
{\large REFERENCES}
\end{centerline}

\noindent
 Demanqul P., Reliannis C.P., Sarajedini A., 1991,
``Observational Tests of Cosmological Inflation" ed. T. Shanks et al.
(Kluwer Academic Publishers), 111

\noindent
 Freedman W.L. et al., 1994 , Nature {\bf 371}, 757

\noindent
 Fukugita M., Hogan C.J., Peebles P.J.E., 1993, Nature {\bf 366}, 309

\noindent
 Jacoby G.H. et al., 1992, PASP {\bf 104}, 599

\noindent
 Hawking S.W., Ellis G.F.R., 1973, {\em The large scale
structure of space-time } (Cambridge University Press, Cambridge),
Chapter 4

\noindent
 Lauer T.R., Postman M., 1992, Astrophys. J. {\bf 400}, L47

\noindent
 Nakamura T. et al., 1995, in preparation

\noindent
 Nakao K., Gouda M., Chiba T., Ikeuchi S., Nakamura T.,
 Shibata M., 1995, Preprint KUNS-1323, OU-TAP-15, YITP/K-1100

\noindent
 Pierce M.J. et al., 1994, Nature {\bf 371}, 385

\noindent
 Renzini A., 1991,
 Observational Tests of Cosmological Inflation, ed. T. Shanks et al.
 (Kluwer Academic Publishers), 131

\noindent
 Smarr L., York J.W., 1978, Phys. Rev. {\bf D17}, 2529

\noindent
 Turner E., Cen R., Ostriker J.P., 1992, AJ {\bf 103}, 1427

\noindent
 van den Bergh S., 1992, PASP {\bf 104}, 861.

\noindent
Wald R.M., 1984, {\it General Relativity}(The University of
Chicago Press, Chicago), Chapter 9


\end{document}